\documentclass[twocolumn,showpacs,preprintnumbers,amsmath,amssymb]{revtex4}
\usepackage{graphicx}
\usepackage{dcolumn}
\usepackage{bm}
\begin{document}
\title{Evolutionary model with genetics, aging and knowledge}
\author{Armando Ticona Bustillos}
\email{aticona@if.uff.br}
\author{Paulo Murilo C. de Oliveira}
\email{pmco@if.uff.br}
\affiliation{Instituto de F\'{\i}sica, Universidade Federal 
Fluminense, 24210-340 Niter\'oi-RJ, Brazil}

\date{\today}

\begin{abstract}
We represent a process of learning by using bit strings, where 1-bits represent the knowledge acquired by individuals. Two ways of learning are considered: individual learning by trial-and-error; and social learning by copying knowledge from other individuals, or from parents in the case of species with parental care. The age-structured bit string allows us to study how knowledge is accumulated during life and its influence over the genetic pool of a population after many generations. We use the Penna model to represent the genetic inheritance of each individual. In order to study how the accumulated knowledge influences the survival process, we include it to help individuals to avoid the various death situations. Modifications in the Verhulst factor do not show any special feature due to its random nature. However, by adding years to life as a function of the accumulated knowledge, we observe an improvement of the survival rates while the genetic fitness of the population becomes worse. In this latter case, knowledge becomes more important in the last years of life where individuals are threatened by diseases. Effects of offspring overprotection and differences between individual and social learning can also be observed. Sexual selection as a function of knowledge shows some effects when fidelity is imposed.
\end{abstract}
\maketitle

\section{Introduction}\label{sec:int}
Learning is any relatively permanent change in behavior which occurs as a result of experience or practice \cite{ref1}. Learning is necessary for an individual to gain autonomy from its natural, physiological and social restrictions, leading the population to improve its life conditions \cite{ref2}. Some theories say that behavior is in part inherited from parents \cite{ref3}, but there is general agreement that the environment is the main factor influencing it. Experiences that produce knowledge cannot be inherited from parents, as is the case for the genome, but rather are acquired in the course of life from different elements of the environment surrounding each individual \cite{ref3}. The process of learning is very complicated and different for each individual of a population. Although the overall conditions for all individuals could be the same, after this process each individual has a different level of knowledge. Differences in the environment or experiences in life, although sometimes very hard to notice (for example: presence of parents during childhood, interaction with other individuals or material environment, education, etc.), produce different levels of knowledge and, as a consequence, a different behavior that depends on the environment surrounding an individual \cite{ref4}.

Genetic evolution modifies the frequency distribution of survival strategies in a population, whereas learning modifies the probability distribution of survival strategies in the repertoire of an individual \cite{ref5}. Reproduction of simple organisms (for example: haploids) does not allow them to improve very much their genetic code, but higher organisms reproduce in a more effective way, in order to escape death (for example: asexual and sexual diploids) \cite{ref6,ref7}. Likewise, primitive organisms present similar behavior and responses to the environment, while higher organisms present a more diverse behavior in order to get a better fitness to the environment \cite{ref2}. This depends on the cognition capacity, i.e. all ways in which animals take in information about the world through the senses, process, retain and decide to act on it \cite{ref8}. However, we clearly know that this knowledge affects the fitness of individuals with their environment, modifying their survival skills and finally improving the survival rate of the whole population. Some learned behavior is first introduced into an individual repertoire and, depending on its consequences, will be acquired by the whole population, modifying even its longevity \cite{ref9}. As it has been observed by Hinton and Nowlan, learning has a drastic effect on evolution \cite{ref10}.

In this paper we present a simple model to study the process of learning using a bit string to represent the knowledge acquired by individuals as a result of interactions with their natural environment (individual learning) and other individuals (social learning). This bit string has an age structure, where each individual has a probability to learn, at every year, from the elements mentioned before. This finally gives as a result the differences in the level of knowledge of individuals. We consider the Penna model to represent the genetic part of individuals (inherited from parents and immutable during one individual's life). Using our model, we study how this genetic part could change after many generations as a consequence of the knowledge acquired by ancestors. These effects have already been studied using another simpler model, first by Hinton and Nowlan \cite{ref10}, and later by Fontanari and Meir \cite{ref7}, showing the importance of learning in evolution.

The paper is organized in the following way: In the first part we explain the Penna model and a preliminary modification to this model, in order to represent families. After that, we explain how knowledge is represented in our model, and the two ways of learning. Then, we present the results of our simulations which were obtained using different values of probabilities, learning strategies and the two ways of relating knowledge to survival. We also consider a case where females are allowed to select males as a function of their knowledge. Finally we discuss our results, their implications and the limitations of our model.

\section{The Penna Model}\label{sec:pen}
The original asexual Penna model \cite{ref11} represents each individual of the population by a bit string of zeros and ones, where each bit position corresponds to a ``year''. In this paper we use the sexual Penna model introduced by Bernardes \cite{ref12,ref13} which corresponds to a reproductive regime of diploid organisms, the population being divided into males and females. These age structured models are very useful to represent and study some important characteristics in the evolution of populations \cite{ref14}.

In this model each individual is represented by two bit strings of size $A_{max}$ read in parallel, where each pair of homologous bits corresponds to one ``year'' in the life of the individual. Genetic diseases are represented by 1 bits in the strings. If an individual has two bits equal to 1 in the same position (homozygote), it will start to suffer the effects of an inherited disease from that year on until its death. There is a limit number $T$ of diseases each individual can accumulate: if at some age an individual has already acquired $T$ diseases, it dies at that time step. In the present work, we do not consider dominant positions, i.e. positions chosen at the beginning of the simulation, where a presence of a bit 1 in one of the strings (heterozygote) represents a disease for the individual. In the study of the fixation of bad genes (explained below) it has been noticed \cite{ref15}, that these dominant positions accumulate 1 bits with less probability, what is called the repulsion effect. This effect does not modify the essence of our results.

An individual dies not only when the number of accumulated diseases reaches the threshold $T$, but also as a result of competition for food and space against other individuals, represented by the logistic Verhulst factor:

\begin{equation}\label{eq1}
V = \frac{N(t)}{N_{max}},
\end{equation}
where $N_{max}$ is the maximum population size the environment can support and $N(t)$ is the current population size. At each time step and for each individual a random number between zero and one is generated and compared with $V$: if it is smaller than $V$, the individual dies independently of its age or genome.

Every time step each female with age equal to or greater than the minimum reproductive age $R$ randomly chooses a male with age also equal to or greater than $R$ to mate, generating $B$ offspring. This mating process is repeated every year until death. The offspring genome is constructed in the following way: the mother genome is cut in a random position, generating four bit string pieces. Two complementary pieces, each one coming from one of the original strings, are joined to form the offspring string which contains the genetic charge to be inherited from the mother. After this, $M$ random mutations are included. Only deleterious mutations are considered: if a 0 bit is tossed to mutate, it is changed to 1; on the other hand if a 1 bit is chosen it remains equal to 1. The same procedure is repeated with the father genome, to produce the second string of the baby. The sex of the newborn is randomly chosen. All this process of testing the survival of each individual and the process of reproduction, both applied over the whole population, represents a time step, i.e. one year in the simulation.

After many generations, stability is reached and the population self-organizes, presenting some properties that can be measured. Stability means that the average number of individuals of any given age is constant in time as well as the concentration of 1-bits at every string position (locus) averaged over the whole population. The first property to be measured is the survival rate \cite{ref16} given by:

\begin{equation}\label{eq2}
S(i) = \frac{n(i+1)}{n(i)},
\end{equation}
where $n(i)$ is the current number of individuals with age $i$. The survival rate represents the probability an individual with age $i$ has to reach age $i+1$.

Another characteristic is the fixation of bad genes (1 bits) \cite{ref15}. Due to the Darwinian dynamics of the model, the population self-organizes allowing the survival of only the best-fitted individuals, i.e. those that present a rather clean genome before the minimum reproduction age: only these individuals live enough to generate offspring and their genomes are passed on to future generations. Because mutations are unavoidable, after many generations they accumulate at the last part of the genome, corresponding to old ages. That is why aging appears \cite{ref11,ref16} (i.e. the continuous decrement of the survival rate) which starts just after the reproduction age. The probability to find individuals with heterozygote positions is high up to the reproduction age, because these positions do not represent any disease for them. On the other hand, homozygote positions, with both bits set to 1, are not usual before the minimum reproduction age, keeping individuals alive until they can reproduce.

\subsection{The Penna Model with Families}\label{sse:pfa}
In order to simulate the teaching process from parents to offspring, we introduce some modifications in the Penna model creating some arrangements representing families. Now a female that reaches age $R$ chooses a male to mate and establishes a family in order to take care of the offspring. Each offspring, when born, has a probability equal to 50\% to stay into the family or to live alone. We consider that the purpose of a family is parental care, so a family without offspring staying in the family, will not continue to be a family: in this case, parents become single again, waiting to search for a new partner in the next time step. Otherwise they continue as a couple breeding again every year, and taking care of the offspring. If one parent dies, the other searches for a new partner to mate and to take care of the offspring. If both parents die, offspring do not keep the family together and live alone from that year on. An offspring that reaches age $R$ should leave the family and look for a partner to mate and try to start its own family. All these conditions mimic characteristics of the behavior of animals that practice parental care \cite{ref17,ref18}.

So far, the only change in the model is to include a concept of fidelity in the process of reproduction which, if violated, does not have any consequence for the offspring, as opposed to the model proposed by Sousa and Moss de Oliveira \cite{ref19}. No explicit advantages are imposed for individuals in or out a family. If a family succeeds to be kept together, in the worst case, it will generate some offspring with similar genomes, like brothers; this would carry an effect in diversity, but our results show that fixation of bad genes presents the same feature as in the normal case. Martins and Penna \cite{ref20} studied a model of infidelity with selection, where off-couple offspring live longer than the ones generated by faithful parents. In their model, a female sometimes prefers an off-couple partner older than her social mate: we did not use this kind of selection in our simulations. In the section corresponding to sexual selection, we will comment on some consequences of fidelity in our model with families, when selection as a function of knowledge is allowed.

Using the same initial condition as in the standard Penna model, the overall characteristics of the population are not affected from the very beginning of the simulation, and at stability averages are the same. The values for survival rate and fixation of bad genes obtained with the standard Penna model are reproduced too, so our modifications did not affect neither the fitness of individuals nor diversity in the population as a whole (In the present work we consider the quantity of genotypes and the frequency of heterozygotes positions, as a measurement of diversity).

\section{The Role of Knowledge}\label{sec:kno}
Now we consider a third bit string, also with the same age structure and the same size $A_{max}$, used to represent the genome. As opposed to genomic bit strings, this third string has all the bits set to 0 at birth, but will change during life. Every time step, an individual has a certain probability to acquire some knowledge at that age. This knowledge will be represented by a 1 bit in that position of the third string, otherwise the 0 bit will be kept, representing no learning at all at that year. Here we do not consider all the processes of learning, which are very complicated and take different periods of time to reinforce what has been learned in order not to be forgotten \cite{ref5,ref8}. This time depends on the cognitive capacity of the individual. We consider an ``acquired knowledge'' only when it has been completely assimilated by the individual, no matter how this occurs.

This age structured model deals with knowledge in such a way that each position represents something which can be learned at that age, like walking and communicating at the beginning of life, self-protecting and getting food supply some time later, etc. Obviously we consider only the kind of knowledge which could be useful for survival \cite{ref2}. Although it would be difficult for the individuals to select what to learn, an individual grows and learns in order to acquire the kind of knowledge which is important to improve its surviving skills \cite{ref2}.

Now, an individual with age $i$ accumulates a quantity $C(i)$, the sum of knowledge bits, which can be used to improve its survival capacity \cite{ref2,ref8,ref9}. As we have seen, in the Penna model an individual dies due to two reasons, accumulation of diseases or the Verhulst factor. The latter represents competition for food or space and depends on a number randomly tossed for each individual at every time step. We model the improvement of survival probability for an individual as a function of the quantity of acquired knowledge $C(i)$ by a new Verhulst factor given by:

\begin{equation}\label{eq3}
V' = V\left[1 - \frac{C(i)}{A_{max}}\right].
\end{equation}

If the quantity $C(i)$ is zero (no acquired knowledge), the old Verhulst factor is kept and the individual dies with the same probability as in the model without knowledge. On the other hand, if a ``$A_{max}$'' years old individual accumulated a quantity $C(A_{max})$ equal to $A_{max}$, its probability to die via Verhulst would be zero. Of course, no individual reaches an age equal to $A_{max}$, due to aging.

Another possibility for knowledge to help an individual's survival is by adding years to life as a function of the accumulated knowledge, which can be expressed as:

\begin{equation}\label{eq4}
Y = f*C(i),
\end{equation}
where $f$ is a factor regulating the equivalence between quantity of knowledge accumulated and further years of life. So, an individual accumulating $T$ diseases, which would therefore die at some genetically determined age $x$, will have $Y$ further years added to its life span according to its knowledge.

\subsection{Ways of Learning}\label{sse:wol}
There are two ways of learning. The first one, ``individual learning'', is due to the interaction of the individual with its natural environment, in a process of trial-and-error \cite{ref9,ref18}: knowledge is acquired from an individual's own experiences, such as avoiding dangers in nature or determining some new food supply \cite{ref9}. So we consider that an individual has a certain probability, at some age, to have this kind of experience and learn from it: this is represented by switching the 0 bit to 1 in that position of its knowledge string. If this individual does not have this kind of experience or did not succeed in learning from it, it will keep the 0 bit in that position. That probability represents the cognitive capacity of individuals and is different for each species \cite{ref8}, this depends on some physiological characteristics \cite{ref21}.

The second way, ``social learning'', is due to the interactions between individuals, so a na\"{\i}ve individual (observer) which spends some time near another one (teacher, in general a conspecific), can learn something just by copying from this partner, i.e. a 1 bit will be set at that position, (obviously, if the teacher does not have the knowledge corresponding to that year, nothing will be learned by the observer, the 0 bit being kept in that position). Here we are considering together the two kinds of social learning, although the most useful for survival would be ``true imitation'', which directly affects the behavior of the na\"{\i}ve individual \cite{ref4,ref9,ref22,ref23,ref24}. Anyway, ``nonimitative social learning'' is also possible, and could be useful for individuals in an unchanging environment, so it would not be rejected. In the present work, we only consider one species in our population, so the teacher is always a conspecific individual. However, there are some cases, in nature, where some animals can learn from other species \cite{ref18}.

It has been observed in some species that family individuals are more likely chosen as teachers \cite{ref22}, so as a special case of social learning we consider the parents as teachers. This can be observed in species that practice parental care. If an offspring is kept into the family, it will be strongly influenced by its parents behavior and will copy it during its stay within the family \cite{ref17,ref18}. Then, if at some year both parents have the knowledge corresponding to that year, the offspring will learn this and a 1 bit will be set at that position. On the opposite, if both parents do not have the knowledge corresponding to that year, the offspring will keep a 0 bit in that position. Finally, if only one of the parents has the knowledge corresponding to that position, we consider that the offspring will acquire that knowledge with a 50\% probability.

\section{Results}\label{sec:res}
We start our simulations using the standard Penna model and with all the positions in the bit strings of the whole population equal to zero. After around 40000 time steps the population has self-organized with the genetic characteristics shown before. Then, from the 50000th time step on, we let the population start to acquire knowledge and wait for another 50000 time steps in order to take averages during the next 20000 time steps. The values used for all the simulations shown in this paper are in Table~\ref{tab:table1}.

\begin{table}
\caption{\label{tab:table1} Values used in the simulations.}
\begin{ruledtabular}
  \begin{tabular}{ll}
Quantity&Value\\
\hline
    Maximum population size & $N_{max} = 10^5$\\
    Size of the bit strings & $A_{max} = 32$\\
    Maximum number of deleterious mutations & $T = 3$\\
    Minimum reproduction age & $R = 10$\\
    Mutation rate & $M =1$\\
    Birth rate & $B = 1$
  \end{tabular}
\end{ruledtabular}
\end{table}

First we study the case of individual learning, considering different probabilities for an individual to learn randomly, simulating the cognitive capacity of individuals. These probabilities are kept constant, the same for the whole population. As a second step we study the social learning allowing interactions between individuals: at every year an individual can learn by itself, or, with a certain probability, it can copy some knowledge from another individual randomly chosen.

Finally we study the effect of families in the process of learning. An individual living without a family follows the same rules of learning explained before. Parents living in families also follow these same rules. On the opposite, offspring living in a family suffer overprotection from their parents, so they can only learn from parents and, with a smaller probability, from other individuals. They have no possibility to acquire experiences by themselves, i.e. they are not allowed to have individual learning.

\subsection{New Verhulst Factor}\label{sse:rvf}
First we consider the case where accumulated knowledge modifies the Verhulst factor for each individual at every year, beginning with the simplest case of individual learning. We used three different probabilities, say $p =$ 25\%, 50\% and 75\%, for an individual to acquire knowledge at every year. It can be seen that the probability of finding an individual with knowledge at any age is the same as the fixed value used in the simulation. The survival rate curve and the fixation of bad genes do not suffer any change, compared to the results obtained for the Penna model without knowledge, in any of the three probabilities used.

\begin{figure}[!ht]
  \begin{center}
    \includegraphics[width=8.6cm]{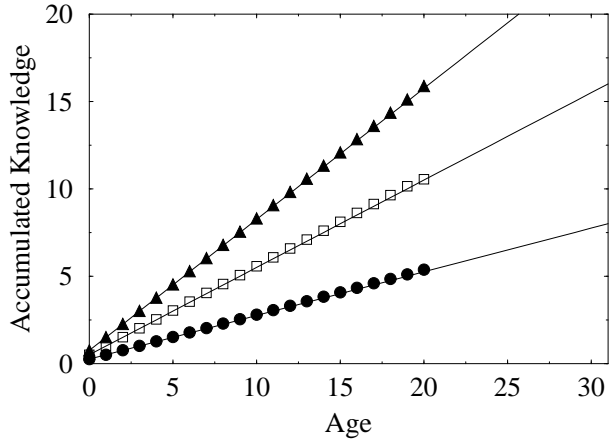}
    \caption{Accumulated knowledge as a function of age, $p=0.25$ (full circles), $p=0.5$ (open squares), $p=0.75$ (full triangles). Linear functions $p*(i +1)$ are also plotted (solid lines).}
    \label{fig1}
  \end{center}
\end{figure}

Figure \ref{fig1} shows the average accumulated knowledge as a function of age. They fit exactly the linear function: $C(i) = p*(i + 1)$. Then, we interpret that in this uninteresting case the process of learning is just accumulated due to the probability we fixed a priori.

\begin{figure}[!ht]
  \begin{center}
    \includegraphics[width=8.6cm]{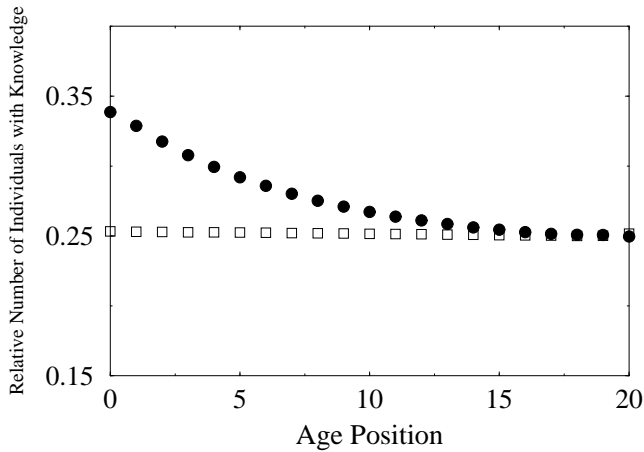}
    \caption{Relative number of individuals randomly acquiring knowledge as a function of age, with $p=0.25$, and copying from others individuals with probability equal to 0.25 (full circles). The former case where individuals learn only from natural environment with $p=0.25$ is also shown (open squares), for comparison.}
    \label{fig2}
  \end{center}
\end{figure}

As a second step we allow social learning due to interactions between individuals. Figure \ref{fig2} shows results for a simulation where each individual has a probability to learn by itself (individual learning) equal to 25\%, and to copy knowledge from others (social learning) with a probability equal to 25\%. For a better understanding, we also plot the curve of knowledge corresponding to the case of individual learning alone, with probability equal to 25\%. Now, the behavior of the acquired knowledge can be understood as a function of the probability of getting knowledge by itself plus the probability of finding an individual that has the knowledge corresponding to that year.

In our simulation we did not put any restriction in choosing a teacher, so the probability of finding a younger individual increases with age. For example if a na\"{\i}ve individual with age equal to 5 chooses as teacher an individual with age equal to 2, therefore without knowledge corresponding to age 5 yet, the na\"{\i}ve individual does not have anything to learn from this teacher. It is easy to see that the probability of finding an individual with acquired knowledge reduces with age and finally for the last year (allowed by aging) there is nobody older with the knowledge corresponding to that year. The figure shows this decreasing probability and at the last point of the curve it has the same value of the curve corresponding to individual learning alone.

Concerning the survival rates and fixation of bad genes, they behave exactly in the same way as in the cases studied before. Then, for this case where knowledge modifies only the Verhulst factor, no special behavior can be noticed and the features observed depend only on the learning probabilities we fixed a priori. We also tested other values of probability and no qualitative change has been observed. In the case of families responsible for the parental care, the behavior of the acquired knowledge is the same as in the former case. Also, the survival rates and the fixation of bad genes do not suffer any change, so we left this case to be treated in the next section.

\subsection{Adding Years to Life}\label{sse:ryl}
Now we consider the case where accumulated knowledge adds years to life. As we mentioned before, in this case we need a conversion factor to relate acquired knowledge with years, so in all our simulations we use the value $f=0.5$; then two 1 bits in the knowledge string represent one more year of life. Other values for this factor do not change the qualitative behavior of our results. Although the chosen value could be considered to be over rated, it helps to give a better idea of the effects we will now explain.

\begin{figure}[!ht]
  \begin{center}
    \includegraphics[width=8.6cm]{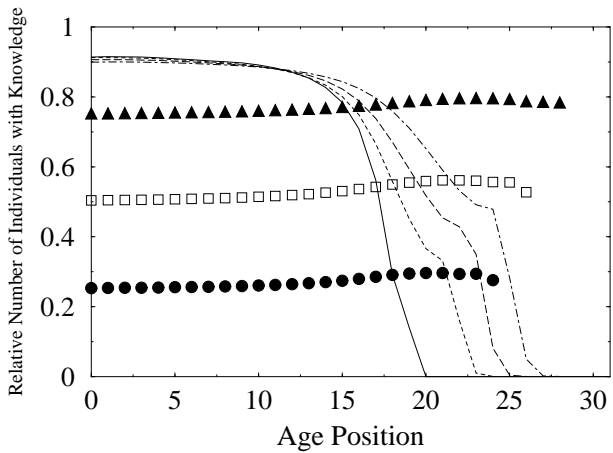}
    \caption{Relative number of individuals with knowledge as a function of the age position, $p=0.25$ (full circles), $p=0.5$ (open squares), $p=0.75$ (full triangles). Survival rate is also plotted for the standard Penna model (solid line), $p=0.25$ (dotted line), $p=0.5$ (dashed line), $p=0.75$ (long dashed line).}
    \label{fig3}
  \end{center}
\end{figure}

We start again with the simplest case of individual learning. Figure \ref{fig3} shows the probability to find an individual with acquired knowledge as a function of age position along the bit string. We use again three values for the probability to acquire knowledge at every year, say 25\%, 50\% and 75\%. For a better understanding, we also plot in the same figure the survival rates for these three cases and the one obtained using the Penna model without knowledge. Now, the survival rate is improved as a function of the probability of acquiring knowledge. Individuals can reach higher ages when they have a higher probability to acquire knowledge, i.e. species with a higher cognitive capacity can better improve their longevity \cite{ref9}.

Another interesting feature is the higher probability of finding individuals with knowledge in the last years, effect that was not observed in the simulation where only the Verhulst factor was modified by knowledge (Figures \ref{fig1} and \ref{fig2}).

The accumulated knowledge as a function of age fits the linear function corresponding to the probabilities fixed in the simulations only during the first ages, where selection keeps the genome clean and the probability to die for this reason is low. This situation changes at older ages because the survival probability is lower, then, the accumulated knowledge comes to compensate diseases in genome. At some ages this effect becomes very strong and tends to stop aging, presenting a knee in the survival rate (Figure \ref{fig3}).

\begin{figure}[!ht]
  \begin{center}
    \includegraphics[width=8.6cm]{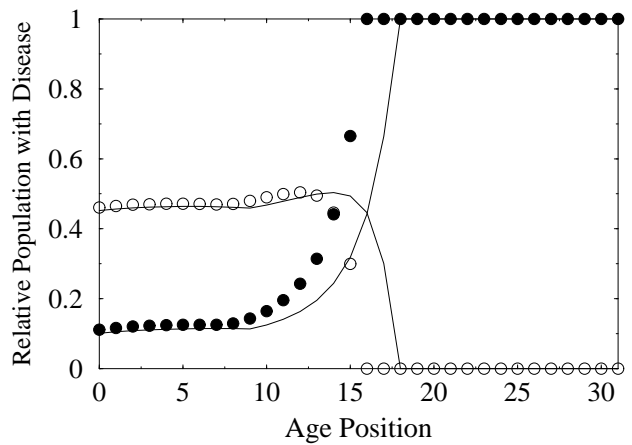}
    \caption{Relative number of individuals: heterozygote positions (standard Penna model solid line, model with knowledge open circles) or homozygote positions (standard Penna model solid line, model with knowledge full circles), as a function of the age position.}
    \label{fig4}
  \end{center}
\end{figure}

As an example, in Figure \ref{fig4} we show the fixation of bad genes for a probability of acquiring knowledge equal to 25\%. We also plot the case for the standard Penna model. Although survival rate has been improved, the genetic fitnesses are now worse than in the normal case, i.e. the probability of finding 1 bits in the positions is slightly greater before the reproduction age, and bad genes accumulate two years earlier, in the last part of the genome. If we used a higher probability of acquiring knowledge, this effect would be stronger. This further supports the statement that knowledge compensates for the accumulation of diseases in the genome. So a population with a high level of knowledge can afford to be less fit to its environment.

A study of the case where social learning is allowed shows these same characteristics improving slightly the overall survival rate, although the genetic fitness of population becomes even worse. As in the case where knowledge modifies the Verhulst factor, we can observe that at early ages the probability of finding an individual with knowledge in these positions is high but this probability reduces for later ages and finally at the last years we observe the same values of probability as in the former case, i.e. individuals with knowledge in the last years are preferred by natural selection.

\begin{figure}[!ht]
  \begin{center}
    \includegraphics[width=8.6cm]{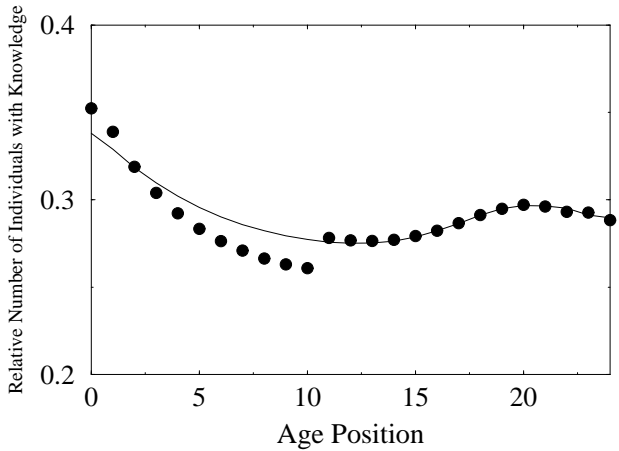}
    \caption{Relative number of individuals with knowledge as a function of the age position; model with families (full circles), model with individual and social learning without families (solid line).}
  \label{fig5}
\end{center}
\end{figure}

As a special case of social learning, we study the case where parents take care of offspring and transmit their knowledge to them. Results for the survival rate and the fixation of bad genes are the same as for the last case, i.e. slightly better for the survival probability and worse for the fitness. In Figure \ref{fig5} we see the probability of finding an individual with knowledge as a function of the age position. We plot together the case with families and the one with individual and social learning without families. The differences at early ages are due to offspring living within families: since they are not allowed to have individual learning, their level of knowledge is conditioned by the one of the teachers. This agrees with the statement that social learning is less adaptive than individual learning \cite{ref9,ref22}. Observers learn only what teachers know, so if the environment changed, the acquired knowledge would not be useful in this new situation.

\subsubsection{Sexual Selection}\label{sse:sel}
Finally we study the case where, at the time of reproduction, females prefer males with high accumulated knowledge \cite{ref22}. We simulated two cases: first, only females with high accumulated knowledge show this mating pattern; second, all females select knowledgeable males. Results are not very different although the last case seems to be slightly better in improving the survival rate. The main consequence of this selection is the improvement of the offspring knowledge due to the higher probability to have a father with acquired knowledge.

\begin{figure}[!ht]
  \begin{center}
    \includegraphics[width=8.6cm]{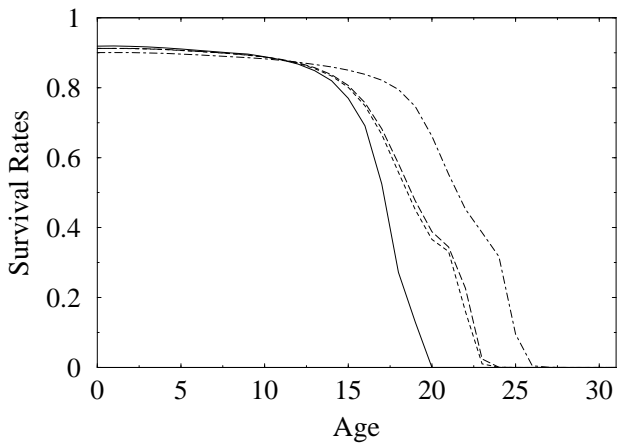}
    \caption{Survival rates for the standard Penna model (solid line), knowledge acquired only from natural environment (dotted line), model with families (dashed line), from natural environment and sexual selection (long-dashed line).}
    \label{fig6}
  \end{center}
\end{figure}

The survival rates of the population are improved, when this kind of sexual selection is allowed. For the case with individual and social learning, this improvement is very noticeable. On the other hand, in the case with families this improvement is hardly noticeable. The difference is due to fidelity because, in the case with families, selection can be done just once, while in the other cases selection is repeated at every time step. So, without fidelity, selection as a function of the level of knowledge is more efficient in improving the survival rates, as can be seen in Figure \ref{fig6}. The fitness is also improved, because the average genome of the population is cleaner than in all the cases considered before, even the case without knowledge.

\section{Discussion}\label{sec:dis}
Although our model for a process of learning seems to be very simple to represent such a complicated process, it reproduces some characteristics of animal behavior as a function of the acquired knowledge. The first way to improve the surviving process by knowledge did not affect at all the characteristics of the standard Penna model due to the random nature of the Verhulst factor, but these results help us in the analysis of the behavior in the other case where knowledge can add years to life.

There are some theories about variation of intelligence during life \cite{ref1,ref21}. Intelligence is defined as the capacity of learning, so we should expect that this capacity would not be a constant through life. We tested some variations in the probability of acquiring knowledge as a function of age (increasing or decreasing this probability as the individual grows), but by modifying only the Verhulst factor, the probability of finding an individual with acquired knowledge follows exactly the shape of the function we used. Also, the survival rate did not change at all. In the case of adding years to life, results follow the same shape of the functions only at the beginning but in the last years this probability is higher. However, these differences were harder to notice due to the shape of the functions. That is why we only used a constant probability of acquiring knowledge. This does not seem to be so far from reality because, according to some authors, intelligence is developed during the early years of life, but then it keeps constant until death \cite{ref21}, although some features in behavior affect the capacity of learning in the last years \cite{ref1}.

In the case of adding years to life as a function of the acquired knowledge, the genetic part becomes worse than in the standard Penna model. This can be understood as the compensation of genetic deficiencies with knowledge, but we can also understand this as an effect of the kind of activity an individual develops during life. So, if an individual spends most of its time in mental activities, in most cases it presents a high level of knowledge, but it becomes physically worse adapted to a natural environment. This will also have inherited consequences in the next generations. This result can be compared with some experimental results of the mortality in the cities and in the country, and also comparations between statistics of countries with differents levels of education and development \cite{ref25,ref26,ref27}.

As we said, our model reproduces the compensation of genetic deficiencies with knowledge, explaining the higher probability to find an individual with knowledge at the last years of life. At the beginning of life this compensation does not seem important, but at older ages when the probability to die is higher due to aging, it becomes more important, so only individuals with knowledge keep alive. This has already been stated by Kaplan and Robson \cite{ref21}, showing the differences between species with a different size of brain, that is directly reflected in the cognitive capacity.

The most important feature our model reproduces is the improvement of survival rates as a consequence of knowledge, a well known fact: the development of a society is measured by its life expectancy. The same can be noticed comparing species with different cognitive capacity \cite{ref21}, where mortality at older ages seems to be smaller for species with more capacity. When we use knowledge to add years to life, we always observe the improvement of survival rates, although the case with sexual selection without families also improves the fitness of the population. The best fitted individuals will also be the ones with more accumulated knowledge, so, after many generations, natural selection and sexual selection performed by females yield a population with only the best fitted individuals and accumulated knowledge.

It has already pointed out, that only considering the genetic part in modeling population dynamics, we are not able to reproduce the behavior of mortality at older ages \cite{ref28}. So other factors have already been considered \cite{ref28,ref29,ref30,ref31,ref32}. In the present work, we tried to add a social factor which in an independent way, has consequences in the survival rates at older ages, of course, this would never be the only factor, but some results show that there is a strong consequence in the mortality due to the knowledge \cite{ref25,ref26,ref27}.

The results about the consequences of parents overprotection seem to reflect also in some way the differences between children who are not under a constant protection of their parents and the ones who are. The first ones sometimes are better prepared to face some difficulties because they have learned by their own experiences to deal with hard situations that have been avoided by the parents of overprotected children. However, we can see here one of the drawbacks of our model: it does not consider that some parents try to offer to their offspring the knowledge they do not have. This could produce some improvements in the process of learning of overprotected children.

Another characteristic reflected in the case with parental care is the difference between social and individual learning. Offspring in families were not allowed to have individual learning, so within this kind of individuals the probability to find acquired knowledge is smaller than for the rest of the population, because it is conditioned by the level of knowledge of the teachers. In nature, the process of learning by trial-and-error seems to be better adaptive to the environment and individuals can respond better to any change in the environment \cite{ref9,ref22}.

\begin{acknowledgments}
We thank J.S. S\'a Martins and D. Stauffer for helpful discussions and critical readings of the manuscript. This work was partially supported by CLAF and Brazilian agencies CAPES, CNPq and FAPERJ.
\end{acknowledgments}

\end{document}